\documentclass[12pt,preprint]{aastex}
\newcommand{\lsim}{\, \lower2truept\hbox{${< \atop\hbox{\raise4truept\hbox{$\sim$}}}$}\,}
\newcommand{\gsim}{\, \lower2truept\hbox{${> \atop\hbox{\raise4truept\hbox{$\sim$}}}$}\,}

\newcommand{\puncspace}{\ifmmode\,\else{\ifcat.\C{\if.\C\else\if,\C\else\if?\C\else%
\if:\C\else\if;\C\else\if-\C\else\if)\C\else\if/\C\else\if]\C\else\if'\C%
\else\space\fi\fi\fi\fi\fi\fi\fi\fi\fi\fi}%
\else\if\empty\C\else\if\space\C\else\space\fi\fi\fi}\fi}
\newcommand{\SP}{\let\\=\empty\futurelet\C\puncspace}

\shorttitle{UV and FIR galaxies at z=0} 

\shortauthors{Xu et al.}

\begin{document}

\title{UV and FIR selected star-forming galaxies at z=0: differences and overlaps}

\author{C. Kevin Xu \altaffilmark{1}, 
Veronique Buat\altaffilmark{11}, 
Jorge Iglesias-P\'{a}ramo\altaffilmark{12},
Tsutomu T. Takeuchi\altaffilmark{13},
Tom A. Barlow\altaffilmark{1},
Luciana Bianchi\altaffilmark{5},
Jose Donas\altaffilmark{2},
Karl Forster\altaffilmark{1},
Peter G. Friedman\altaffilmark{1},
Timothy M. Heckman\altaffilmark{7},
Patrick N. Jelinsky\altaffilmark{8},
Young-Wook Lee\altaffilmark{6},
Barry F. Madore\altaffilmark{9,10},
Roger F. Malina\altaffilmark{2},
D. Christopher Martin\altaffilmark{1},
Bruno Milliard\altaffilmark{2},
Patrick Morrissey\altaffilmark{1},
R. Michael Rich\altaffilmark{3},
Susan G. Neff\altaffilmark{4},
David Schiminovich\altaffilmark{1},
Oswald H. W. Siegmund\altaffilmark{8},
Todd Small\altaffilmark{1},
Alex S. Szalay\altaffilmark{7},
Barry Y. Welsh\altaffilmark{8},
Ted K. Wyder\altaffilmark{1}, and
Sukyoung Yi\altaffilmark{6}
}

\altaffiltext{1}{California Institute of Technology, MC 405-47, 1200 East
California Boulevard, Pasadena, CA 91125}
\altaffiltext{2}{Laboratoire d'Astrophysique de Marseille, BP 8, Traverse
du Siphon, 13376 Marseille Cedex 12, France}
\altaffiltext{3}{Department of Physics and Astronomy, University of
California, Los Angeles, CA 90095}
\altaffiltext{4}{Laboratory for Astronomy and Solar Physics, NASA Goddard
Space Flight Center, Greenbelt, MD 20771}
\altaffiltext{5}{Center for Astrophysical Sciences, The Johns Hopkins
University, 3400 N. Charles St., Baltimore, MD 21218}
\altaffiltext{6}{Center for Space Astrophysics, Yonsei University, Seoul
120-749, Korea}
\altaffiltext{7}{Department of Physics and Astronomy, The Johns Hopkins
University, Homewood Campus, Baltimore, MD 21218}
\altaffiltext{8}{Space Sciences Laboratory, University of California at
Berkeley, 601 Campbell Hall, Berkeley, CA 94720}
\altaffiltext{9}{Observatories of the Carnegie Institution of Washington,
813 Santa Barbara St., Pasadena, CA 91101}
\altaffiltext{10}{NASA/IPAC Extragalactic Database, California Institute
of Technology, Mail Code 100-22, 770 S. Wilson Ave., Pasadena, CA 91125}
\altaffiltext{11}{
Observatoire Astronomique Marseille Provence, Laboratoire 
d'Astrophysique de Marseille, 13012 Marseille, France}
\altaffiltext{12}{
Instituto de Astrofisica de Andlucia (CSIC), Camino Bajo de Huetor 50,
18008 Granada, Spain}
\altaffiltext{13}{
Astronomical Institute, Tohoku University, Aoba, Aramaki, Aoba-ku, Sendai
980-8578, Japan}

\begin{abstract}

  We study two samples of local galaxies, one is UV (GALEX) selected
  and the other FIR (IRAS) selected, to address the question whether
  UV and FIR surveys see the two sides ('bright' and 'dark') of the
  star formation of the same population of galaxies or two different
  populations of star forming galaxies. No significant difference
  between the L$_{tot}$ ($=L_{60}+L_{FUV}$) luminosity functions of
  the UV and FIR samples is found.  Also, after the correction for the
  `Malmquist bias' (bias for flux limited samples), the FIR-to-UV
  ratio v.s. L$_{tot}$ relations of the two samples are consistent
  with each other.  In the range of $9 \la \log(L_{tot}/L_\sun) \la
  12$, both can be approximated by a simple linear relation of $\log
  (L_{60}/L_{FUV})=\log(L_{tot}/L_\sun)-9.66$.  These are consistent
  with the hypothesis that the two samples represent the same
  population of star forming galaxies, and their well documented
  differences in L$_{tot}$ and in FIR-to-UV ratio are due only to the
  selection effect. A comparison between the UV luminosity functions
  shows marginal evidence for a population of faint UV galaxies
  missing in the FIR selected sample.  The contribution from these
  'FIR-quiet' galaxies to the overall UV population is 
  insignificant, given that the K-band luminosity functions (i.e. the
  stellar mass functions) of the two samples do not show any
  significant difference.

\end{abstract}


\section{Introduction}

The evolution of star forming galaxies tells much about the history of
the universe. The star formation activity in these galaxies can be
best studied by observing the emission from young massive stars in the
rest frame UV and FIR. The UV observations record the direct light
from the hot young stars, and the FIR observations collect star light
absorbed and then re-emitted by the ubiquitous dust. A complete
picture of star formation in the universe can only be obtained when
the observations in these two wavebands are properly synthesized.
Indeed, our knowledge on the star formation history of the universe
 has been mostly derived from deep surveys in the
rest frame UV and FIR. Many studies have been devoted to methods of
deriving star formation rate of individual galaxies using the UV or
FIR luminosities (Calzetti 1997; Meurer et al. 1999; Buat \& Xu
1996; Iglesias-P\'{a}ramo et al. 2005),
and the strengths and shortcoming of these methods have been discussed
thoroughly in the literature (Kennicutt 1998; Adelberger \& Steidel
2000; Bell 2002; Bell 2003; Buat et al. 2005; Kong et al. 2004; 
Iglesias-P\'{a}ramo 
et al. 2006).  However, an arguably more important issue is the
selection effect of the surveys that can be summed up by the following
question: Do UV and FIR surveys see the two sides ('bright' and 'dark')
of the star formation of the same population of galaxies, or do they see
two different populations of star forming galaxies?  This is important
because if the correct answer is the latter, then even
if one can estimate accurately the star formation rate for galaxies in
surveys in one band, the star formation in galaxies detected
in the other band is still missing. 
Actually this question is in the core of an
on-going debate on whether the SFR of z$\sim 3$ universe can be
derived from observations of Lyman-break galaxies, which are UV
selected star forming galaxies at z$\sim 3$ (Adelberger and Steidel
2000), given that SCUBA surveys in sub-millimeter (rest frame FIR for
z$\ga 2$) detected many violent star forming galaxies at about the
same redshift that are not seen by LBG surveys (Smail et al. 2001;
Smail et al. 2004).

There have been limited overlaps between rest frame UV surveys
and rest frame IR surveys.  In the SCUBA survey of LBG's (Chapman et
al. 2000), only one LBG was detected. As summarized in Adelberger and
Steidel (2000), only a couple of SCUBA sources are bright enough in
optical to be detected in LBG surveys.  The situation is better for
z$\sim 1 $ star forming galaxies, which now can be routinely
identified by large scale spectroscopic surveys and multi-band optical
surveys, and which have been detected in abundance in mid-infrared 
by ISOCAM deep surveys (Elbaz et al. 2002; Hammer et al. 2005) and
Spitzer surveys (Le Floc\'h 2005), and
in UV by GALEX (Arnouts et al. 2005; Burgarella et al. 2006).
However the extrapolation from the mid-IR to
the total dust emission is very uncertain and may subject to
significant evolution itself.
The same criticism can also be
applied to the comparisons between rest frame UV and MIR sources at
z$\sim 2$, the latter being detected recently by Spizter at
24$\micron$ (Chary et al. 2004; Shupe et al. 2005). 
Because of the relatively high confusion limits
for surveys in
the Spitzer MIPS 70$\mu m$ and 160$\mu m$ bands, thorough comparisons
of rest frame UV and FIR sources of z$\ga 1$, down to luminosity
levels fainter than the 'knee' of the luminosity functions of both
bands, may have to wait until the launch of Herschel (Pilbratt 2005).

In this paper, we investigate the difference and overlaps of the UV
and FIR selected samples in the local universe in an attempt to shed
light on the selection effects of high-z samples similarly selected.
The UV data are taken from observations by Galaxy
Evolution Explorer (GALEX) and the FIR data are taken from the IRAS
database. Several papers have been published using these data. Martin
et al. (2005, hereafter M05) derived the local (z=0) bivariate luminosity function
for the FUV (1530{\AA}) and FIR (60$\mu m$) bands, which shows that
the FUV luminosity saturates at about 2 10$^{10}$ L$_\sun$ while the FIR
luminosity can be as high as $\sim 10^{13}$ L$_\sun$.  This is
consistent with a very strong dependence of the FIR/FUV ratio on the
total luminosity ($\rm L_{tot}=L_{FIR} + L_{FUV}$).  The luminosity
function of $L_{tot}$ has a log-normal form.  Buat et al. (2005)
compared the extinction properties of local UV and FIR selected
galaxies and found that the mean NUV (2267{\AA}) extinction of UV
selected galaxies is significantly lower than that of FIR selected
galaxies ($\sim 1$ mag vs. $\sim 2.5$ mag).  Iglesias-P\'{a}ramo et al. (2006)
carried out an extensive study, using combined GALEX and IRAS data, on
the UV and FIR emission as star formation indicators and found a
rather modest star formation activity for local star forming
galaxies. Pre-GALEX studies on comparisons between UV and FIR selected
samples can be found in Buat \& Burgarella (1998), Buat et al. (1999)
and Iglesias-P\'{a}ramo et al. (2004).
In this work, we will study some statistics free of the 
selection effect in order to check quantitatively how much the UV
and FIR selected samples differ/overlap with each other.
The paper is arranged as following: After this introduction,
the data sets analyzed in this paper are presented in Section 2.
Major results are listed in Section 3. Section 4 is devoted to the
discussion. Through out this
paper, we assume $\Omega_\Lambda = 0.7$, $\Omega_m = 0.3$, and H$_0$ =
70 km sec$^{-1}$ Mpc$^{-1}$.

\section{Data}
The data sets are basically the same as those in Iglesias-P\'{a}ramo et
al. (2006) and Buat et al. (2005).  The original UV selected sample
(Iglesias-P\'{a}ramo et al. 2006) includes 95 galaxies brighter than NUV =
16 mag selected from GALEX G1 stage All-sky Imaging Survey (AIS),
covering 654 deg$^2$.  From these we exclude 1 galaxy,
2MASX\_J20341333-0405, which does not have measured redshift.  The FIR
selected sample is also taken from Iglesias-P\'{a}ramo et al. (2006).  From
the original sample, including 163 galaxies with $f_{60} \geq
0.6$ Jy in 509 deg$^2$ sky covered both by GALEX AIS and IRAS PSCz
(Saunders et al. 2000), 2 are excluded: NGC~7725 (no redshift) and
IRAS-F00443+1038 (not a galaxy).  Consequently, the UV and FIR
selected samples studied in this paper have 94 and 161 galaxies,
respectively. The ${\rm K_s}$ (2.16$\mu m$) band magnitudes $\rm
K_{tot}$ are taken from Extended Source Catalog (XSC) of 2MASS (Jarrett
et al. 2000). Since this is very close to the classical K (2.2$\mu m$)
magnitude, we will call it K magnitude hereafter for the sake of
simplicity. For both the UV selected sample (94 galaxies) and
the FIR selected sample (161 galaxies), each has
12 galaxies undetected by 2MASS. According to the sensitivity limit of
2MASS XSC (Jarrett et al. 2000), 
upper limits of K=13.5 mag are assigned to the undetections.

\begin{figure}
\plotone{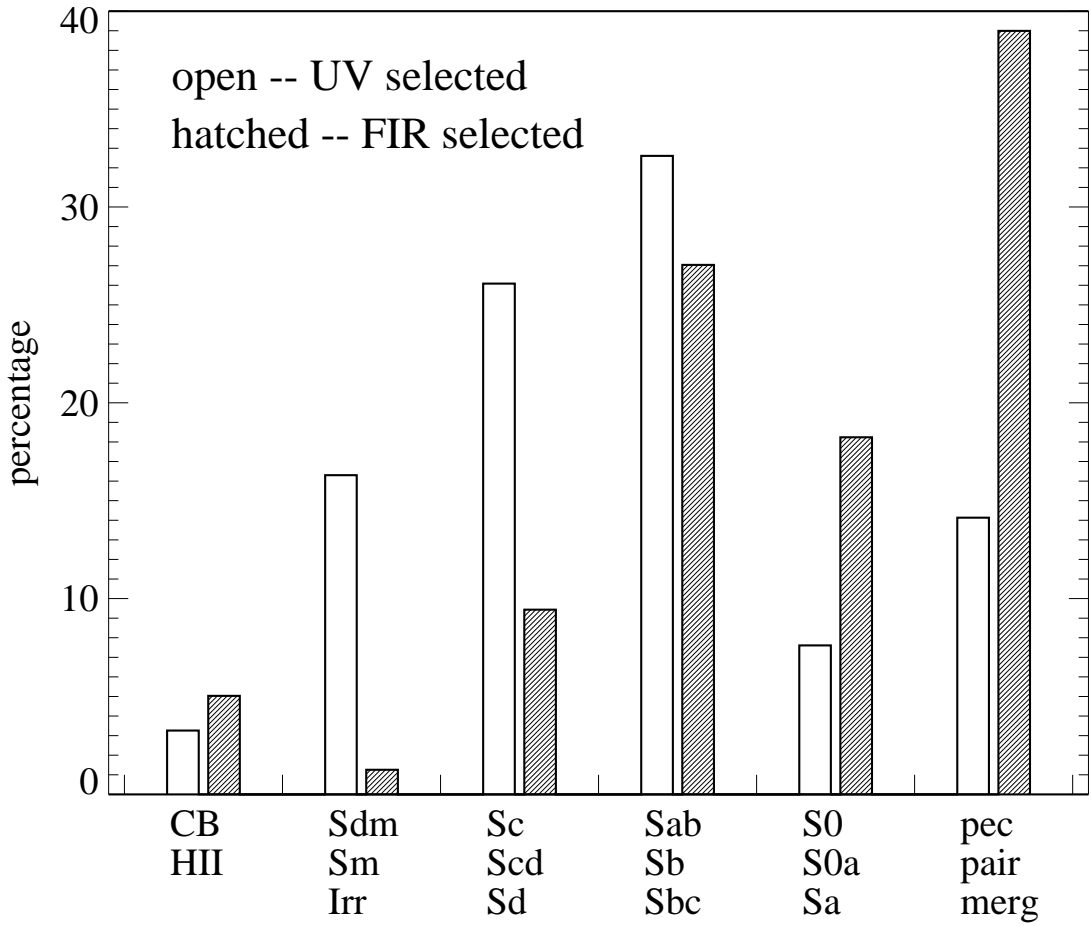}
\caption{Morphological type distributions 
of the UV and FIR selected samples.} 
\end{figure}

Morphological classifications were searched in NED.  For
those galaxies without morphological classification in the literature,
images taken from SDSS, DSS and 2MASS (in the order of priority) were
inspected and eye-ball classification was carried out. Galaxies included
in Atlas of Peculiar Galaxies (Arp 1966), Southern Peculiar Galaxies
and Associations (Arp \& Madore 1987), and Catalog of Isolated Pairs
of Galaxies (Karachentsev 1972) are classified as peculiar,
interacting or mergers.  For a few galaxies that are faint (b $> 15$
mag) and small ($\la 10''$) the classification can be very uncertain.
Most of such galaxies are in the FIR selected sample, and very often
there is clear sign of interaction (close companion of similar
brightness and/or diffuse tidal features). 
In Fig.1 the distributions of morphological types (not including QSOs
and ellipticals) of the two samples are compared. The overall
overlap between the two distributions is about 60\%. There is a
significant excess of Pec/Int/merg galaxies in the FIR selected
sample (39\%) compared to those in the UV selected sample (14\%). For
normal galaxies both UV and FIR selected samples peak in the bin of
Sab/Sb/Sbc. Detailed analysis shows that the median type for normal UV
galaxies is Sc and that of normal FIR galaxies is Sb. 
The FIR selected sample is tilted toward the earlier spirals 
whereas the UV sample has more late type (later than Sc) galaxies. 

\section{Results}
\subsection{Comparisons of luminosity functions of UV and FIR galaxies}
Much of the difference between UV and FIR selected samples 
can be traced back to a single
selection effect: UV observations detect preferentially galaxies
with low L$_{FIR}/L_{UV}$ ratios, and in contrast FIR observations 
select galaxies with high L$_{FIR}$/L$_{UV}$ ratios.
Since the FIR/UV ratio is a good indicator
of dust attenuation (Xu \& Buat 1995; Buat \& Xu 1996;
Meurer et al. 1999; Gordon 2000), it follows that
UV samples select galaxies with significantly lower
dust attenuation than galaxies in the FIR selected sample:
Buat et al. (2005) found a median FUV attenuation of 
$A(FUV) =0.8^{+0.3}_{-0.3}$ mag
for the UV selected sample, compared to a
$A(FUV) =2.1^{+1.1}_{-0.9}$ mag for the FIR selected sample.

It has been well established that there is a strong correlation
between luminosity and dust attenuation in the sense that
more luminous galaxies have higher dust attenuation
(Wang \& Heckman 1996; Buat \& Burgarella 1998; Adelberger \& Steidel 2000;
M05). Fig.2 shows that galaxies in both
FIR and UV samples follow the strong $\rm L_{60}/L_{UV}$
vs. L$_{tot}$ correlation. On the other hand,
 UV galaxies in general have significantly 
lower L$_{tot}$ and lower  $\rm L_{60}/L_{UV}$ ratios for a given
L$_{tot}$ compared to FIR galaxies (Buat \& Burgarella 1998;
 Adelberger \& Steidel 2000;
Iglesias-P\'{a}ramo et al. 2006). Can these trends be 
attributed solely to the selection effects, or do they reflect
some intrinsic differences between the two populations?
\begin{figure}
\plotone{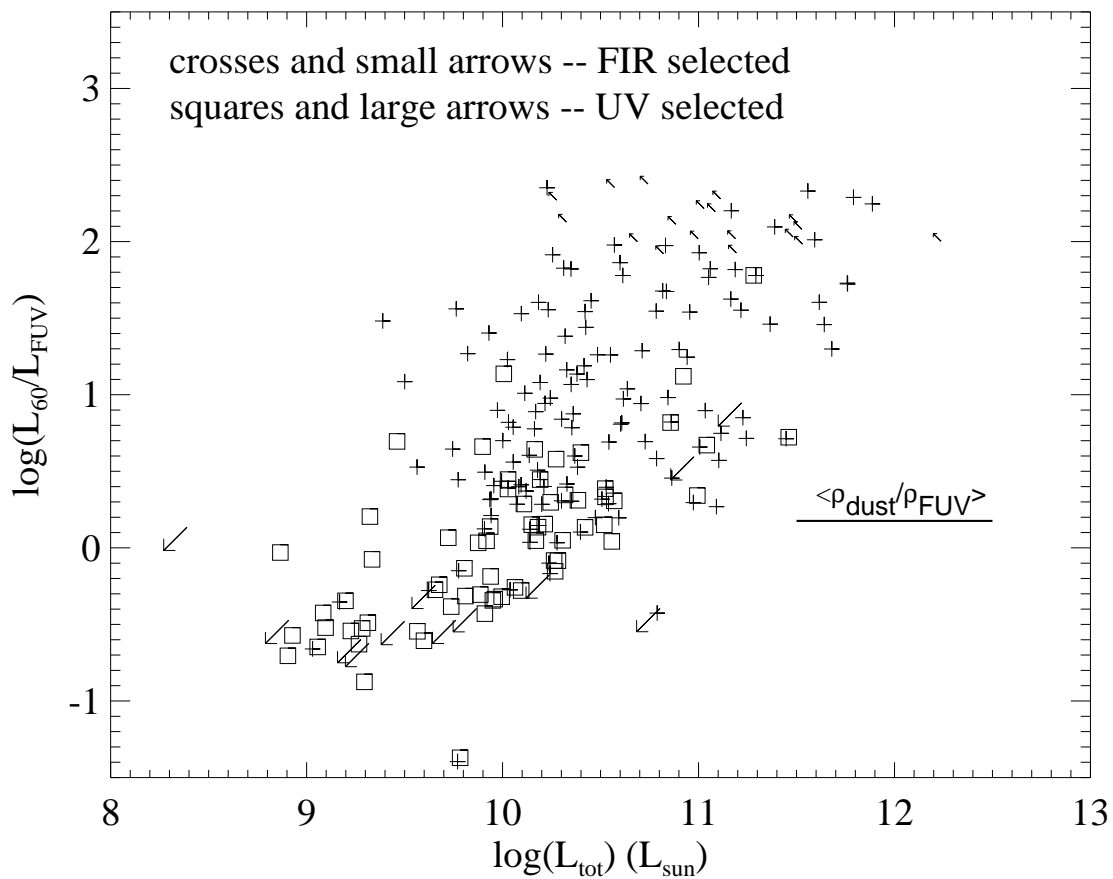}
\caption{The L$_{60}$/L$_{FUV}$ ratio vs. 
L$_{tot}$ (L$_{60}$+L$_{FUV}$) plot for
UV (blue symbols) and FIR (red symbols) selected galaxies.
The cosmic mean of the FIR/UV ratio, $<\rho_{dust}/\rho_{FUV}>$
is taken from Takeuchi et al. (2005), assuming L$_{60}=0.4\times$ L$_{dust}$.}
\end{figure}

In order to answer this question, we have to compare the 
statistics of L$_{tot}$ and of $\rm L_{60}/L_{UV}$ that are free
from the selection effect. The selection effect
is introduced by the so-called `Malmquist bias' 
on both flux limited samples:  For a given L$_{tot}$,
galaxies with higher FIR-to-UV ratios
have brighter $L_{60}$, therefore can be seen at
larger distances (i.e. having a larger 
maximum finding volume V$_{max}$) in a $f_{60}$ limited sample. 
Similarly, for a given L$_{tot}$,
galaxies with lower FIR-to-UV ratios have higher
$L_{UV}$ and therefore larger 
$V_{max}$ in a UV flux limited sample. In what follows
we shall compare the L$_{tot}$ luminosity functions
(LFs hereafter)
of the two samples to exam whether they have the same intrinsic 
L$_{tot}$ distributions. Because LFs are 
luminosity distributions
of galaxies in a unit volume, they are not subject to the
bias discussed above.

Here
we exclude the sources whose IRAS fluxes are affected by the cirrus.
Also UV galaxies not covered by the IRAS survey are dropped.
This reduces the FIR sample to 151 galaxies and the UV sample to 81
galaxies. The IRAS detections of 5 UV galaxies are confused with other
UV sources, therefore the corresponding IRAS fluxes are treated as upperlimits.
Altogether 14 UV galaxies have only upperlimits for the IRAS flux.
For galaxies in the FIR selected sample, 14 have only upperlimits 
for the FUV flux. 

Define $\rm \phi_{tot}^{FUV} (L_k)$ as the L$_{tot}$ LF
of UV selected galaxies at $\rm \log (L_{tot}) = L_k$,
$\rm \phi_{FUV} (L_i)$ 
the FUV (1530{\AA}) LF at $\rm \log (L_{FUV})=  L_i$,
and $\rm P_{k,i}$ the conditional 
probability of finding UV galaxies of $\rm \log (L_{FUV})=L_i$ in 
the bin of $\rm L_k-0.5 \delta_k < \log (L_{tot}) \leq  L_k+0.5 \delta_k$.
Then
\begin{equation}
\rm \phi_{tot}^{FUV} (L_k) 
          =\sum_{i} P_{k,i} \phi_{FUV} (L_i) \delta_i/\delta_k .
\end{equation}
Similarly, the L$_{tot}$ luminosity function of FIR selected sample 
can be derived using the formula:
\begin{equation}
\rm \phi_{tot}^{60} (L_{k'}) 
     =\sum_{j} P_{k',j} \phi_{60} (L_j) \delta_j/\delta_{k'}
\end{equation}
where $\rm P_{k',j}$ the conditional 
probability of finding FIR galaxies of $\rm \log L_{60}=L_j$ in 
the bin of $\rm L_{k'}-0.5 \delta_{k'} < \log (L_{tot}) 
\leq  L_{k'}+0.5 \delta_{k'}$.
Data in our two samples are used in the calculations of the
conditional probability functions $\rm P_{k,i}$ and  $\rm P_{k',j}$.
In order to take into
account the information content in the upper limits,
the Kaplan-Meier (KM) estimator (Kaplan \& Meier 1958; Feigelson \& Nelson
1985; Schmitt 1985) has been applied in these calculations.
We have chosen 
$\delta_i$=1 mag for the L$_{FUV}$ bin width,
$\delta_j$=0.5 dex for the L$_{60}$ bin width, 
and $\delta_k$=0.5 dex for the L$_{tot}$ bin width. Other choices of the
bin widths result in LFs with either larger scatters (bin widths too
narrow) or coarse resolutions (bin widths too broad). The FUV LF and 
L$_{60}$ LF are taken from Wyder et al. (2005) and Takeuchi et al. (2003),
respectively.

\begin{figure}
\plotone{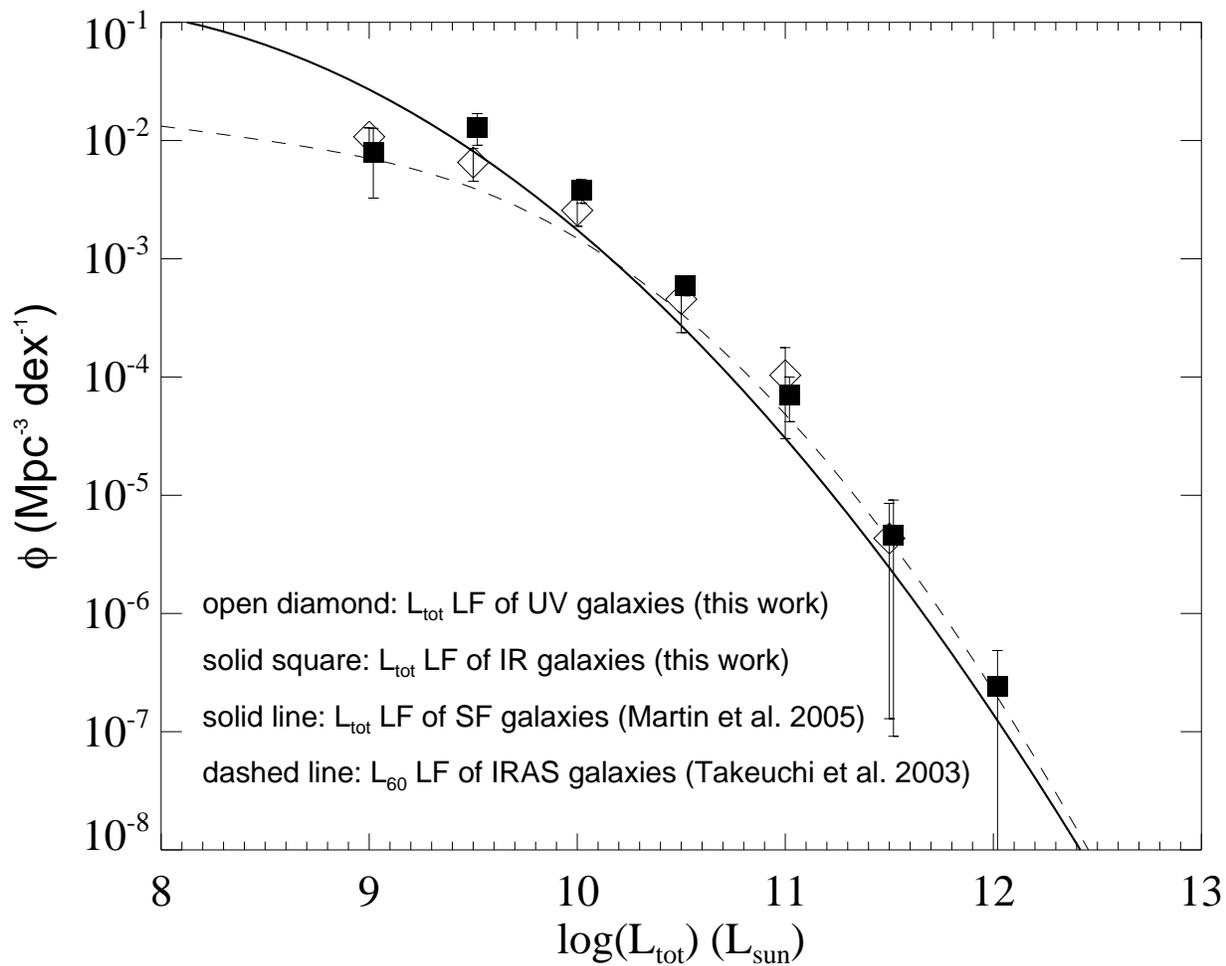}
\caption{The L$_{tot}$ ($\rm L_{60}+L_{FUV}$)
luminosity functions of UV galaxies (open diamonds)
and FIR galaxies (solid squares). 
}
\end{figure}

\begin{deluxetable}{ccccc}
\tablewidth{0pt}
\tablecaption{The  L$_{tot}$ ($\rm L_{60}+L_{FUV}$)
Luminosity Functions of UV and FIR Selected Galaxies.}
\tablehead{\colhead{(1)} & \colhead{(2)} & \colhead{(3)}
 & \colhead{(4)} & \colhead{(5)}
 \\
\colhead{log(L$_{tot}$)} 
& \colhead{$\phi^{FUV}_{tot}$} & \colhead{error} 
& \colhead{$\phi^{60}_{tot}$} & \colhead{error} 
\\ 
\colhead{(L$_\sun$)} &
 \colhead{(${\rm Mpc^{-3} dex^{-1}}$)} & \colhead{(${\rm Mpc^{-3} dex^{-1}}$)} &
 \colhead{(${\rm Mpc^{-3} dex^{-1}}$)} & \colhead{(${\rm Mpc^{-3} dex^{-1}}$)} 
 }
\startdata
       9.0 &  1.076E-2 &   2.039E-3  &  7.948E-3 &  4.739E-3 \\
       9.5 &  6.522E-3 &   2.024E-3  &  1.298E-2 &  3.907E-3 \\
      10.0 &  2.564E-3 &   6.875E-4  &  3.804E-3 &  8.727E-4 \\
      10.5 &  4.548E-4 &   2.200E-4  &  5.978E-4 &  1.124E-4 \\
      11.0 &  1.033E-4 &   7.394E-5  &  7.077E-5 &  2.915E-5 \\
      11.5 &  4.310E-6 &   4.223E-6  &  4.574E-6 &  4.528e-6 \\
      12.0 &   ...     &    ...      &  2.430E-7 &  2.430e-7 \\ 
\enddata
\end{deluxetable}

The results are listed in Table 1 and plotted in Fig.3.
In the L$_{tot}$ range where they overlap,
the LFs of the two populations are consistent with each other.
The solid line is the best fit of the L$_{tot}$ LF of M05,
derived
from a combined sample of UV and FIR selected galaxies. It is
a log-normal function with the center at $\log (L_{tot}/L_\sun) = 7.43$ and
$\sigma = 0.87$. In bins of  $\log (L_{tot}/L_\sun) \ga 10$, our LFs are
marginally higher than that of M05. In order to
check whether this indicates over-estimation in our results,
we also compared with the $\rm L_{60}$ LF of Takeuchi et al. (2003).
There is a good agreement between our results and that of 
 Takeuchi et al. (2003) for bins of  $\log (L_{tot}/L_\sun) \ga 11$
(where  $\rm L_{60}$ always dominates  L$_{tot}$), both are slightly
higher than that of M05. At 
$\log (L_{tot}/L_\sun) = 9$, our results for both samples are lower than that
of M05, possibly due to uncertainties caused by
the small size of our samples compared to that of M05.

The above result is consistent with that the UV and the FIR samples
are drown from the same population of star-forming galaxies.  
However, the L$_{tot}$ LF comparison could be insensitive to some
differences. For example, in bins where L$_{tot}$ is dominated by
L$_{60}$, the differences between L$_{FUV}$ distributions of two
samples can be hidden by the similarity between L$_{60}$ distributions,
and vise versa.  Therefore, in what follows we shall calculate the
L$_{60}$ LF of UV galaxies and compare it with the L$_{60}$ LF of
IRAS galaxies (Takeuchi et al. 2003), and calculate the L$_{FUV}$
(1530{\AA}) LF of FIR galaxies and compare it with that of GALEX galaxies
(Wyder et al. 2005).

The formalism for the calculations of L$_{60}$ LF of UV galaxies,  
$\rm \phi_{60}^{FUV} (L_{60})$, and of L$_{FUV}$ LF of FIR galaxies,  
$\rm \phi_{FUV}^{60} (L_{FUV})$, is the same as that used in the
calculations of L$_{tot}$ LFs. One only needs to replace $L_{tot}$ by
L$_{60}$ in Eq(1), and by L$_{FUV}$ in Eq(2).

\begin{figure}
\plotone{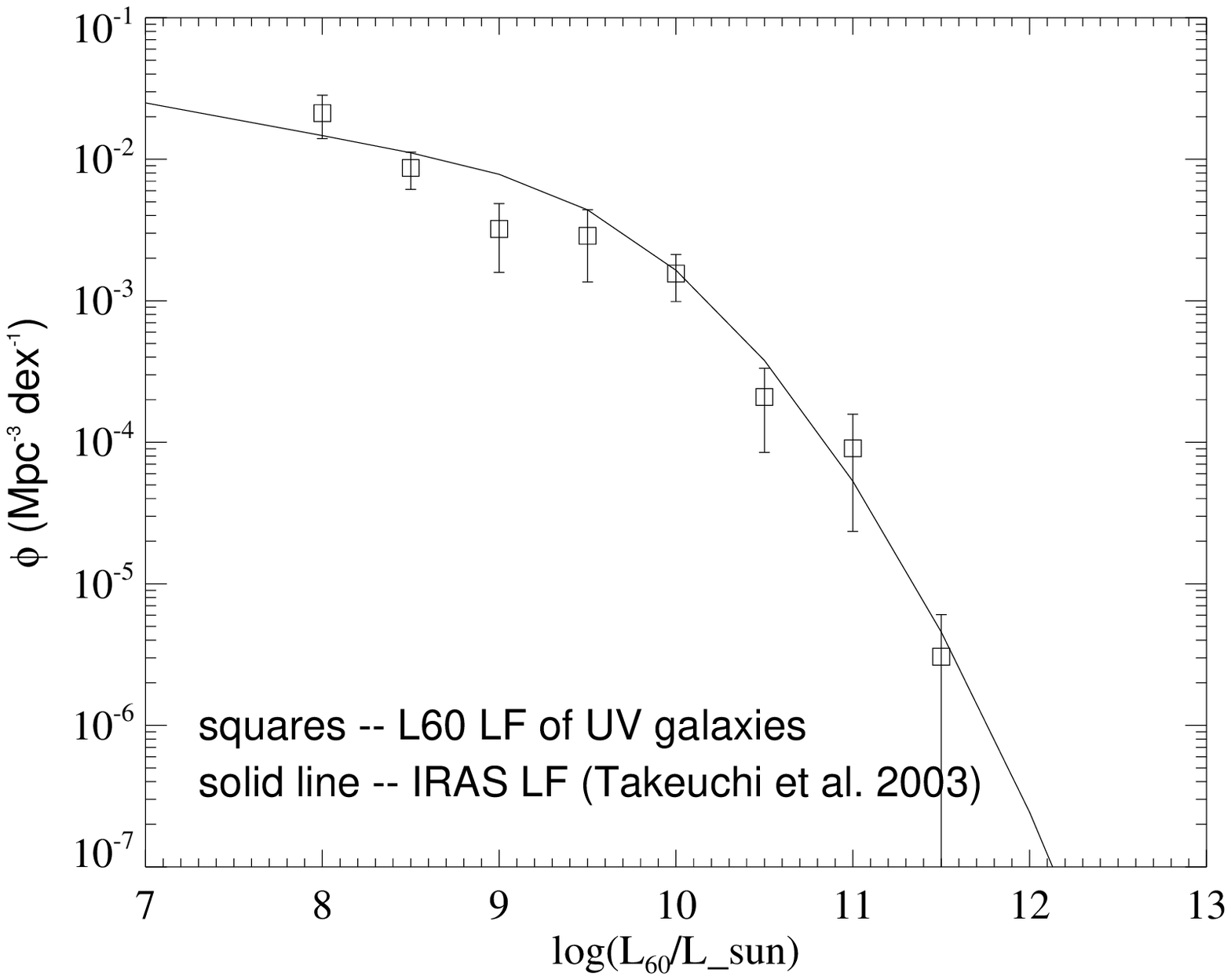}
\caption{The L$_{FIR}$ (60$\mu m$)
luminosity function of UV selected galaxies compared to the
IRAS 60$\mu m$ luminosity function (Takeuchi et al. 2003).}
\end{figure}

\begin{figure}
\plotone{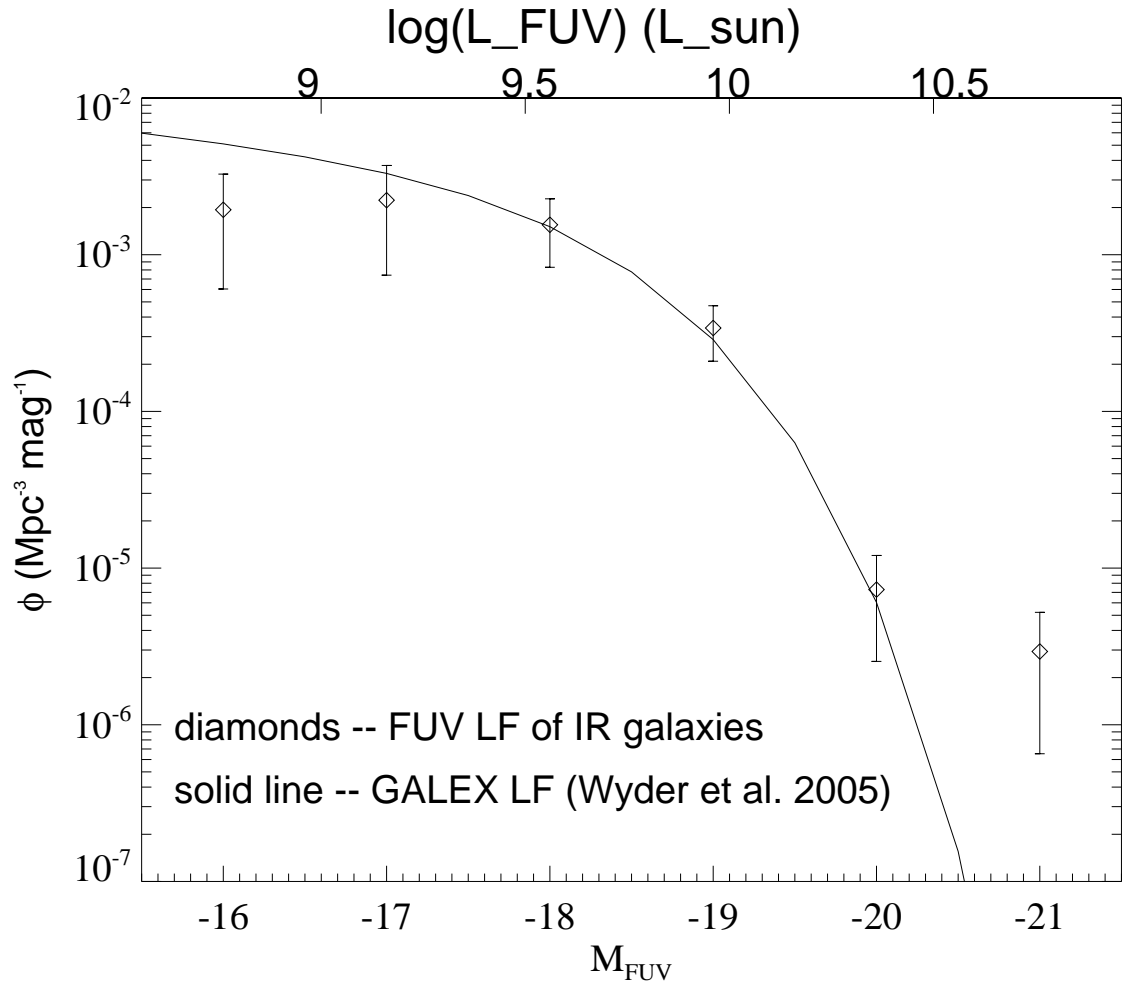}
\caption{The FUV (1530{\AA})
luminosity function of FIR selected galaxies compared to the
GALEX FUV luminosity function (Wyder et al. 2005).}
\end{figure}
\begin{deluxetable}{ccc}
\tablewidth{0pt}
\tablecaption{The FIR (60$\mu m$)
Luminosity Function of UV Selected Galaxies.}
\tablehead{\colhead{(1)} & \colhead{(2)} & \colhead{(3)} \\
\colhead{log(L$_{60}$)} & \colhead{$\phi^{FUV}_{60}$}
& \colhead{error} \\ 
\colhead{(L$_\sun$)} & \colhead{(${\rm Mpc^{-3} dex^{-1}}$)} & \colhead{(${\rm Mpc^{-3} dex^{-1}}$)} }
\startdata
   8.0 & 2.117E-2 &  7.181E-3 \\
   8.5 & 8.679E-3 &  2.548E-3 \\
   9.0 & 3.219E-3 &  1.634E-3 \\
   9.5 & 2.875E-3 &  1.517E-3 \\
  10.0 & 1.555E-3 &  5.680E-4 \\
  10.5 & 2.090E-4 &  1.243E-4 \\
  11.0 & 9.078E-5 &  6.729E-5 \\
  11.5 & 3.057E-6 &  2.996E-6 \\
\enddata
\end{deluxetable}

\begin{deluxetable}{ccc}
\tablewidth{0pt}
\tablecaption{The FUV (1530{\AA})
Luminosity Function of FIR Selected Galaxies.}
\tablehead{\colhead{(1)} & \colhead{(2)} & \colhead{(3)} \\
\colhead{M$_{FUV}$} & \colhead{$\phi_{FUV}^{60}$}
& \colhead{error} \\ 
\colhead{(mag)} & \colhead{(${\rm Mpc^{-3} mag^{-1}}$)} & 
\colhead{(${\rm Mpc^{-3} mag^{-1}}$)} }
\startdata
 -16 & 1.936E-3  &  1.332E-3 \\
 -17 & 2.226E-3  &  1.486E-3 \\
 -18 & 1.552E-3  &  7.211E-4 \\
 -19 & 3.407E-4  &  1.316E-4 \\
 -20 & 7.286E-6  &  4.748E-6 \\
 -21 & 2.934E-6  &  2.281E-6 \\
\enddata
\end{deluxetable}

The results are listed in Table 2 and Table 3.  
In Fig.4, $\phi_{60}^{FUV} (L_j)$ is compared with the 60$\mu m$ luminosity
function of IRAS sources (Takeuchi et al. 2003).  It appears that UV
galaxies can account for the FIR luminosity function up-to L$_{60}\sim
10^{11.5}$ L$_\sun$. Only ULIRGs of L$_{60}\gsim 10^{12}$ L$_\sun$ are
missing in the UV sample. This is because ULIRGs are
very rare in the local universe, and they are much fainter in UV.
Therefore they are probed by UV
surveys in a very much smaller volume compared to that probed
by the FIR surveys.  It should be pointed out that the UV LF of
Wyder et al. (2005) excludes the
contribution from broad-line AGNs identified using SDSS spectra.
These are UV/optical selected QSOs and
Seyfert 1 galaxies. According to Sanders et al. (1989) and Spinoglio
\& Malkan (1989), these sources never contribute more than 10\% of the
IR LF in the whole range of FIR luminosity.  The comparison between FUV
luminosity function of the FIR selected sample and the GALEX FUV
luminosity function (Wyder et al. 2005) is in Fig.5. It shows that UV
galaxies brighter than L$_*$(FUV) ($\sim 10^{9.5}$ L$_\sun$) are fully
represented in the FIR selected sample. In fact there is a significant
excess in the brightest bin ($\rm M_{FUV} = -21$) of the UV LF of FIR
sources compared to the UV LF of Wyder et al. (2005), likely being
caused by the exclusion of the broad-line AGNs in the latter. There
is a marginal evidence for fainter UV galaxies of
 L$_{FUV} < 10^{9.5}$ L$_\sun$ being
under-represented in the FIR selected sample, suggesting that a
population of `FIR-quiet' UV galaxies might be missing in the FIR selected
sample.

\subsection{FIR-to-UV v.s. L$_{tot}$ relations of UV and FIR galaxies}

Let $\rm R = \log (L_{60}/L_{FUV}) = \log (L_{60}) - \log (L_{FUV})$.
For UV galaxies with a given $\rm L_{tot}=L_k$, a
`Malmquist-bias-free' (i.e. selection effect free)
indicator of mean FIR-to-UV ratio can
be defined as follows:
\begin{equation}
\rm R_{UV}(L_k) ={\sum_{j,i} (L_j-L_i) P_{j,i} \phi_{FUV} (L_i) \delta_i
           \over \sum_{j,i} P_{j,i} \phi_{FUV} (L_i) \delta_i}
\end{equation}
where $\rm P_{j,i}$ is the conditional 
probability of finding UV galaxies of $\rm \log (L_{FUV})=L_i$ in 
the FIR luminosity bin 
$\rm L_j-0.5 \delta_j < \log (L_{60}) \leq  L_j+0.5 \delta_j$,
and the summation goes through both indexes i and j including all bins 
satisfying the condition 
$\rm L_k-0.5 \delta_k < \log (10^{L_i}+10^{L_j}) \leq  L_k+0.5 \delta_k$.
A similar FIR-to-UV ratio indicator can be defined for FIR galaxies:
\begin{equation}
\rm R_{FIR}(L_k) ={\sum_{i,j} (L_j-L_i) P_{i,j} \phi_{60} (L_j) \delta_j
           \over \sum_{i,j} P_{i,j} \phi_{60} (L_j) \delta_j}
\end{equation}
where $\rm P_{i,j}$ is the conditional 
probability of finding FIR galaxies of $\rm \log (L_{60})=L_j$ in 
the FUV luminosity bin 
$\rm L_i-0.5 \delta_i < \log (L_{FUV}) \leq  L_i+0.5 \delta_i$.
The variance of $\rm R_{UV}(L_k)$ and that of $\rm R_{FIR}(L_k)$,
respectively, are:
\begin{equation}
\rm \sigma^2_{UV}(L_k) ={\sum_{j,i} [(L_j-L_i)- R_{UV}(L_k)]^2 P_{j,i}\phi_{FUV} (L_i) \delta_i
           \over \sum_{j,i} P_{j,i} \phi_{FUV} (L_i) \delta_i},
\end{equation}
and
\begin{equation}
\rm \sigma^2_{FIR}(L_k) ={\sum_{j,i} [(L_j-L_i)- R_{FIR}(L_k)]^2
            P_{i,j} \phi_{60} (L_j) \delta_j
           \over \sum_{j,i} P_{i,j} \phi_{60} (L_j) \delta_j}.
\end{equation}

\begin{figure}
\plotone{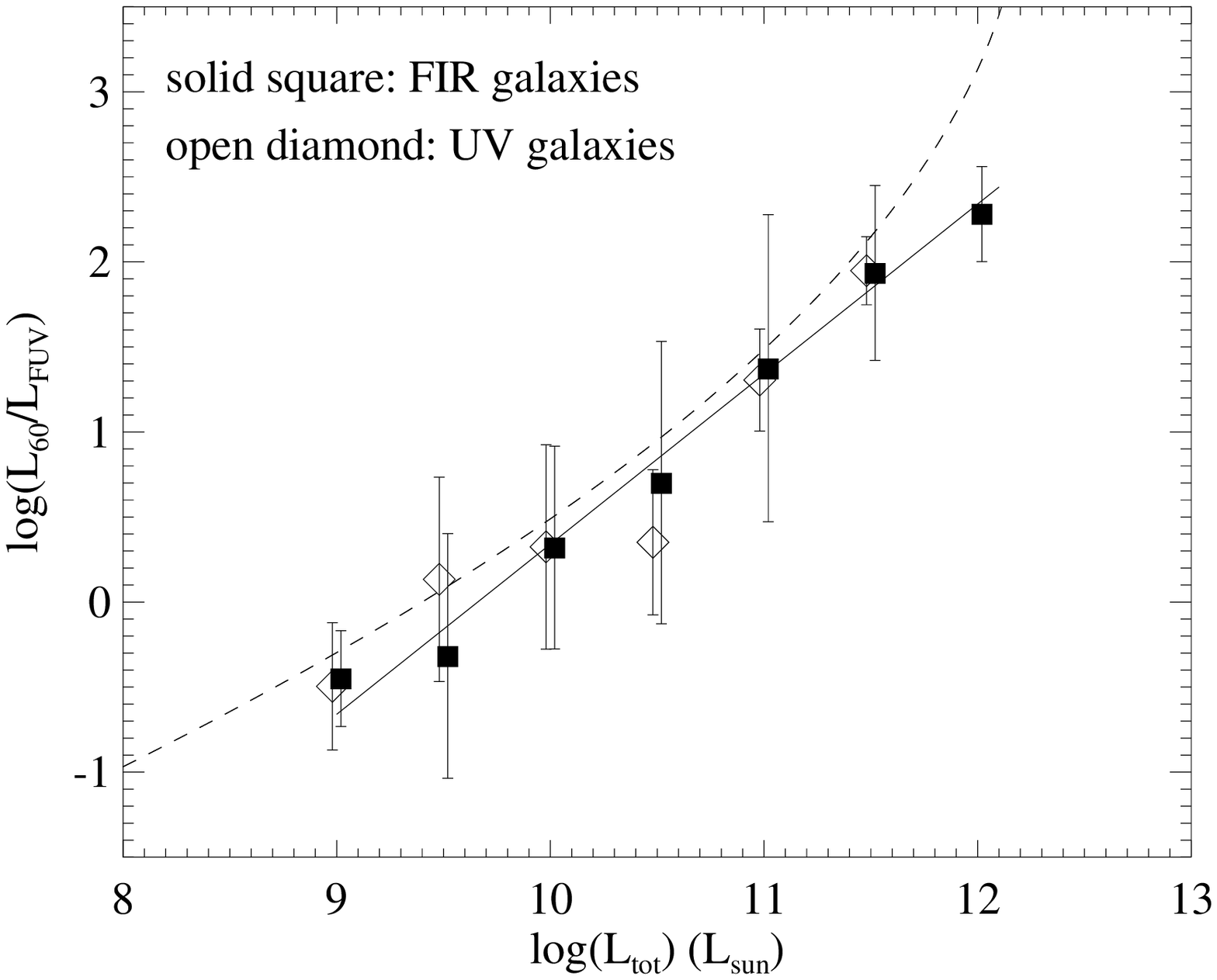}
\caption{
The FIR-to-UV ratio ($R=\log (L_{60}/L_{FUV})$)
v.s. L$_{tot}$ ($\rm L_{60}+L_{FUV}$) relations for UV and FIR Selected Galaxies.
Solid line:  $\rm R = \log (L_{tot}) - 9.66$.
Dashed curve: R v.s. $\rm \log (L_{tot})$ relation
derived by Martin et al (2005) for a combined sample.
}
\end{figure}

\begin{deluxetable}{ccccc}
\tablewidth{0pt}
\tablecaption{The FIR-to-UV ratio ($R=\log (L_{60}/L_{FUV})$)
v.s. L$_{tot}$ ($\rm L_{60}+L_{FUV}$) relations for UV and FIR Selected Galaxies.}
\tablehead{\colhead{(1)} & \colhead{(2)} & \colhead{(3)}
 & \colhead{(4)} & \colhead{(5)} \\
\colhead{log(L$_{tot}$/L$_\sun$)} 
& \colhead{~~$\rm R_{UV}$~~} & \colhead{~~$\sigma_{UV}$~~} 
& \colhead{~~$\rm R_{FIR}$~~} & \colhead{~~$\sigma_{FIR}$~~} 
 }
\startdata
       9.0 &  -0.495  &   0.374   &  -0.451  &  0.282 \\
       9.5 &   0.133  &   0.600   &  -0.320  &  0.722 \\
      10.0 &   0.324  &   0.600   &   0.317  &  0.598 \\
      10.5 &   0.351  &   0.427   &   0.697  &  0.833 \\
      11.0 &   1.305  &   0.299   &   1.370  &  0.907 \\
      11.5 &   1.948  &   0.201   &   1.932  &  0.516 \\
      12.0 &  ...     &    ...    &   2.279  &  0.281 \\ 
\enddata
\end{deluxetable}

Results for $\rm R_{UV}$, $\rm R_{FIR}$, $\rm \sigma_{UV}$,
and $\rm \sigma_{FIR}$, are listed in Table 4. As shown in Fig.6,
there is no significant difference between the
$\rm R_{UV}$ v.s. L$_{tot}$ relation of UV galaxies
and the $\rm R_{FIR}$ v.s. L$_{tot}$ relation of FIR galaxies, 
again in consistence with
the hypothesis that the two samples represent
the same population, and their difference
in Fig2. is due to the selection effect. Both 
$\rm R$ v.s. L$_{tot}$ relations can be approximated by
a simple linear relation: $\rm R = \log (L_{tot}) - 9.66$, as shown
by solid line in Fig.6. In the $\rm L_{tot}$ range covered
by our samples, this relation is slightly lower than the
non-linear relation between R and $\rm \log (L_{tot})$
(dashed curve in Fig.6) derived by M05 from the bi-variate
function of their combined sample. It should be pointed out
that the simple linear relation should not be extrapolated to
galaxies of  $\rm L_{tot} \la 10^9 L_\sun$, where 
a flatter relation is more likely (M05).

\subsection{K band luminosity functions and stellar mass distributions}
%

\begin{figure}
\plotone{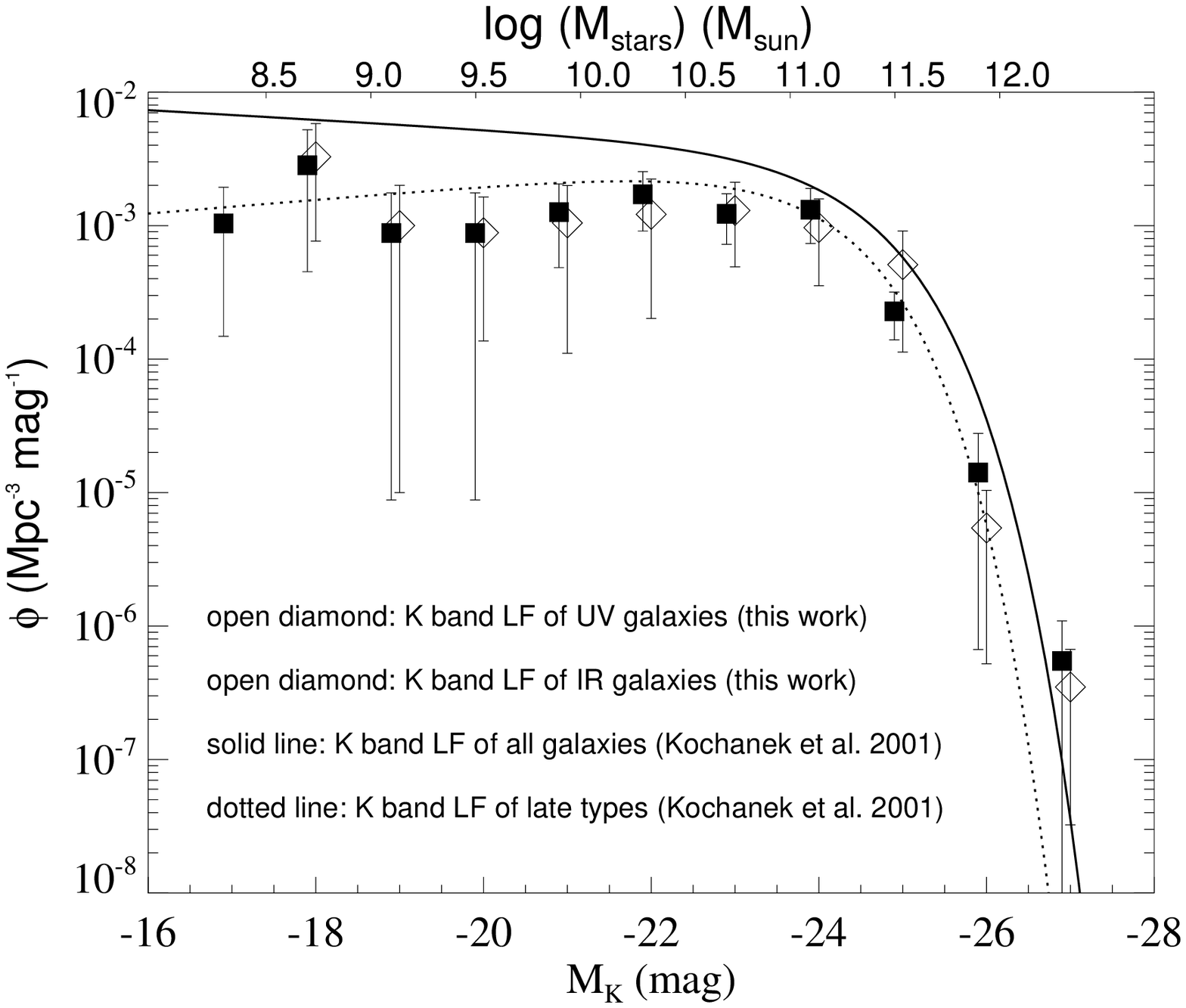}
\caption{K band luminosity functions
(stellar mass distributions) of UV and FIR selected samples.}
\end{figure}

The NIR K band luminosity, very insensitive to both the dust
extinction and the star formation history variation (Bell \& De Jong
1991; Bell et al. 2003), 
is the best stellar mass indicator. The stellar mass distribution
is one of the most important characteristics defining 
galaxy populations, therefore we would
compare the K band LF of UV galaxies with
that of FIR galaxies. Because of the present of upper limits in the
K band fluxes in both UV and FIR samples, we exploit the
same formalism as presented in Eq(1) and Eq(2), 
and use KM estimator in calculating the conditional
probability functions $\rm P(M-0.5\delta <M_K \leq M+0.5\delta | L_{FUV})$
and $\rm P(M-0.5\delta <M_K \leq M+0.5\delta | L_{60})$.
In Fig.7, the resulted K LFs of the two samples are compared with each
other. No significant difference is found between them.
It is interesting to note that both K LFs are consistent
with the K LF of late-type galaxies derived by Kochanek
et al (2001), specified by a Schechter function with 
$\phi_0=0.0101$, $\alpha=-0.87$ and 
$\rm M_* = -22.98+5.*alog10(h_0)-\delta$, where $h_0=0.7$ and
 $\delta=0.2$ (the difference between the isophotal magnitude and
the 'total' magnitude, Cole et al. 2001).
The conversion factor $\rm M_{stars}/L_K =
1.32 M_\sun/L_\sun$, which is derived for a stellar population 
with constant star formation rate and a Salpeter IMF (Cole et
al. 2001), is assumed when converting the
K band luminosity to stellar mass.

\section{Discussion}
Our results indicate that bulk of z=0 galaxies 
selected in the UV and FIR samples
are from the same population of active star forming
galaxies. In particular, galaxies in the two samples
have indistinguishable L$_{tot}$ LFs. And their FIR-to-UV
ratio v.s. L$_{tot}$ relations, after correction for the Malmquist bias, 
are consistent with each other. Therefore, the well documented
results that galaxies in the UV flux limited samples tend to have lower L$_{tot}$
and lower FIR-to-UV ratios for a given  L$_{tot}$ than those galaxies in 
the FIR flux limited samples are purely due to the selection effect.

The only sign for a possible difference between UV and FIR populations
is a marginal deficiency of galaxies of low UV luminosity in the
FIR selected sample, indicating the existence of an 'FIR-quiet' UV
population.  Indeed, it has been known 
that there is a population of low-metallicity, low
dust content 'blue compact dwarf' galaxies. 
The prototype is I~ZW~18, the galaxy with
one of the lowest metallicity of 1/50 solar (Searle \& Sargent
1972). I~ZW~18 has never been detected in FIR. The FUV magnitude of
I~Zw~18 derived from its GALEX image is 15.75 mag (Gil De Paz, private
communication).  Its IRAS upper limit of f$_{60\mu m} < 0.2$
Jy corresponds to a upperlimit of $\rm L_{60}/L_{FUV} < 0.27$.
There are only a few percent of galaxies in our UV sample have
such low  $\rm L_{60}/L_{FUV}$ ratio, indicating a
low contribution from these 'FIR-quiet' galaxies to the 
overall UV population. This is in agreement with the result in Fig.7 which 
shows no significant difference in the K LFs 
(i.e. stellar mass functions) of the UV and FIR galaxies.
Furthermore, because they have rather low 
UV and FIR luminosities, these galaxies contribute negligibly to
the total SFR of the local universe. It will be interesting to know
whether in the earlier universe more star forming galaxies
becoming `FIR quiet', given the lower metallicity in
high z galaxies and marginal evidence for 
net increase of the faint end of the UV LF (Arnouts et al. 2005).
The new results of Burgarella et al. (2006) 
on LBG galaxies at z$\sim 1$ suggest the existence of 
a population of low-attenuation, bright UV galaxies at that redshift.

There are no ULIRGs in our UV sample. It is generally true that 
ULIRGs are absent in UV samples of sizes less than a few 1000s.
In the local universe, LIRGS/ULIRGs contribute less than a few percent
to the total star formation in
all galaxies (Soifer \& Nuegebauer 1998). Therefore the
absence of them in UV selected samples does not
introduce significant bias in the estimate of total starformation
rate. But in the earlier universe
of $z \ga 1$, this bias may be more significant. According to
Le Floc'h et al. (2005), about more than 10\% of star formation at
$z \sim 1$ is due to ULIRGs.

\vskip0.2truecm
\noindent{\it Acknowledgments}:
GALEX (Galaxy Evolution Explorer) is a NASA Small Explorer, launched
in April 2003.  We gratefully acknowledge NASA's support for
construction, operation, and science analysis for the GALEX mission,
developed in cooperation with the Centre National d'Etudes Spatiales
of France and the Korean Ministry of Science and Technology.
We thank an anonymous referee for very constructive comments.


\end{document}